\documentclass[a4paper,10pt]{article}
\usepackage{amsfonts}
\usepackage{amsmath}
\usepackage{amssymb}
\usepackage{graphicx}
\usepackage{caption}

\begin{document}

\title{Valley properties of doped graphene in a magnetic field}
\author{J.S. Ardenghi$^{\dag }$\thanks{%
email:\ jsardenghi@gmail.com, fax number:\ +54-291-4595142}, P. Bechthold$%
^{\dag }$, E. Gonzalez$^{\dag }$, P. Jasen$^{\dag }$ and A. Juan$^{\dag }$ \\
%EndAName
$^{\dag }$IFISUR, Departamento de F\'{\i}sica (UNS-CONICET)\\
Avenida Alem 1253, Bah\'{\i}a Blanca, Buenos Aires, Argentina}
\maketitle

\begin{abstract}
The aim of this work is to describe the electronic properties of graphene in
a constant magnetic field in the long wavelength approximation with random
binary disorder, by solving the Soven equation self-consistently. Density of
state contributions for different valleys in each sublattice sites are
obtained for different values of magnetic field strength showing remarkable
differences between $K$ and $K^{\prime }$ valleys. A band gap is obtained by
an asymmetric on-site impurity concentration and the graphene electrons
acquire an anomalous magnetic moment, which is opposite in different
valleys, which depend highly in the interplay between the impurity band, the
band edges and the broadening of the Landau levels. In turn, magnetization
as a function of $B$ for different on-site random impurities is computed
showing that by decreasing the on-site impurity energy values, maximum
magnetization is shifted towards higher values of $B$ which can be used to
create and manipulate polarized valley currents. Finally, conductivity and
local vertex function are obtained as a function of energy showing that
scattering contributions from $A$ and $B$ sublattices differ significantly.
Effective medium local two-irreducible vertex is computed showing that
scattering from sublattice $A$ and $B$ do not contribute equally, which can
be related to weak anti-localization. From these results, it could be
possible to explore how the valley pseudospin can be used to create
polarized currents by populating asymmetrically the sublattice sites, where
the population can be tuned with the applied magnetic field strength.
\end{abstract}

\section{Introduction}

Graphene is a two-dimensional allotrope of carbon which has become one of
the most significant topics in solid state physics due to the large number
of applications (\cite{novo},\cite{intro1},\cite{intro2}, \cite{B}, \cite%
{BBBB}). The carbon atoms form a honey-comb lattice made of two
interpenetrating triangular sublattices, $A$ and $B$. A special feature of
the graphene band structure is the linear dispersion at the Dirac points
which are dictated by the $\pi $ and $\pi ^{\prime }$ bands that form
conical valleys touching at the two independent high symmetry points at the
corner of the Brillouin zone, the so called valley pseudospin \cite{A}. In
the absence of defects, electrons near these symmetry points behave as
massless relativistic Dirac fermions with an effective Dirac-Weyl
Hamiltonian \cite{B}. When a magnetic field is applied perpendicular to the
graphene sheet, a discretization of the energy levels is obtained, the so
called Landau levels \cite{kuru}. These quantized energy levels still appear
also for relativistic electrons, just their dependence on field and
quantization parameter is different. In a conventional non-relativistic
electron gas, Landau quantization produces equidistant energy levels, which
is due to the parabolic dispersion law of free electrons. In graphene, the
electrons have relativistic dispersion law, which strongly modifies the
Landau quantization of the energy and the position of the levels. In
particular, these levels are not equidistant as occurs in a conventional
non-relativistic electron gas in a magnetic field. This large gap allows one
to observe the quantum Hall effect in graphene, even at room temperature 
\cite{E}. In turn, the valley pseudospin can be used to create polarized
currents in a similar way as the real electron spin is used in spintronics.
By applying a local gate voltage to a quantum nano-contact in a graphene
nanoribbon, valley filter can be obtained. Due to the finite size of the
graphene sheet, the transvesal momentum component are quantized. Those
states with a definite group velocity, depending on the\ polarized gate
voltage sign applied, can go through the quantum nano-contact with the
result of a largest population in one of the valleys. Altough the
theoretical approach to distinguish carriers in the two valleys are known,
the experimental procedure has become an attracting literature (see \cite%
{ryz}, \cite{re}, \cite{xiao} and \cite{bula}). This valley pseudopin
carries real magnetic moment, like real spin, supported by chiral orbital
current, which can rise Zeeman splitting and Pauli paramagnetism in graphene 
\cite{foldy}. This magnetic moment can be used to couple it with a
perpendicular magnetic field, which can be about $30~$times the Bohr
magneton for low energy electrons \cite{ryz}. However, taking into account
the total contributions of the valley pseudospin, there is no net
polarization by the magnetic field because the contribution of the Zeeman
term to the Hamiltonian has opposite sign at the $K$ and the $K^{\prime }$
points, which implies that the direction of the valley pseudospin
polarization induced by the magnetic field cancel each other \cite{ken}. In
this case, the valley degeneracy is protected by spatial inversion symmetry 
\cite{taill}. Besides the usual spatial translation and rotation symmetries,
pristine graphene Hamiltonian is invariant under a large number of
symmetries in the isospin spaces \cite{os}. By introducing the most general
disorder Hamiltonian $H_{dis}=\underset{i,j}{\overset{}{\sum }}V_{ij}\sigma
_{i}\tau _{j}$, pseudospin symmetries can be broken depending on which
elements of $V_{ij}$ are non-zero (\cite{moro} and \cite{falko}). In
particular, local diagonal disorder can be obtained experimentally by
irradiation \cite{esque}, where carbon atoms are extracted from the graphene
plane and adatoms or adsorbed species attach to the graphene plane \cite%
{schedin}, or in which some carbon atoms are chemically substituted for
other elements. These adsorbed particles can induce a local potential at the
sites where they couple to the carbon lattice and change the on-site energy
in the Hamiltonian. For weak random substitutional disorder potential,
intravalley mixing within either the $K$ and $K^{\prime }$ valleys is
possible, but not intervalley mixing. In the case of a random binary alloy
disorder, a gap can be open in the Fermi level due to the broken $C_{z}$
chiral symmetry \cite{os}. The peculiar behavior of the $n=0$ Landau level
in graphene, where its amplitude is nonzero only in one of the sublattices,
namely, at $B$ sites for $K$ valley and $A$ sites for $K^{\prime }$ valley,
combined with the random diagonal disorder can enhance the asymmetry in the
valley amplitudes, which is related to the geometric nature of the Bloch
band and its relation with Berry phase (see \cite{chang}, \cite{xiao2}, \cite%
{thon}, \cite{cere} and \cite{xiao3}). This asymmetry can be detected by a
population difference in the two valleys as a signal of orbital
magnetization (see \cite{xiao}) and in turn, this asymmetry can be enhanced
by an asymmetry in the relative density of states for both valley
pseudospins. In this sense, the aim of this work is to study the behavior of
the valley pseudospin under a constant magnetic field and random binary
disorder. In particular, self-energies and density of states for both
sublattices and both valleys will be obtained by applying coherent potential
approximation (CPA) (see \cite{soven}, \cite{soven2} and \cite{veli}). In
this method, the system is replaced by an effective medium with a complex
self-energy that replace the disorder random potential. The value of the
effective medium self-energy can be obtained by demanding that the average
scattering of an electron by the surrounding medium is zero. Improvement of
the method for better understanding of non local disorder can be done by
replacing the single impurity site by a cluster of atoms (see \cite{tsu}, 
\cite{bis} and \cite{duca}). Due to the unbroken translational symmetry, the
embedded cluster method has been applied (\cite{gon}, \cite{my} and \cite{yu}%
). To obey all the imposed criteria for the CPA (see \cite{gon2}), the Non
Local Coherent Potential approximation has been developed \cite{jar} which
is based in the approximation of continuous lattice functions $f(k)$ by
cluster function $f(K_{n})$, where $K_{n}$ are points in momentum space that
satisfy Born-von Karman boundary conditions. Without taking into account the
improvements of the CPA method, we will consider the results for low energy
Bloch electrons with self-energies in the full Born approximation with
nested diagrams included.\footnote{%
See diagramatic\ techniques for CPA \cite{elliot}.} This work will be
organized as follow: In section II, the magnetic Green function with
diagonal on-site energies will be computed for graphene. Although this
procedure has been studied without diagonal on-site energies, for a
self-contained work this results will be generalized. In section III,
single-site approximation will be applied and a system of coupled Soven
equation will be found and solved. The discussion of the results is shown in
section IV and the principal findings of this paper are highlighted in the
conclusion.

\section{Green function}

For a self-contained lecture of this paper, a brief introduction of the
quantum mechanics of graphene in a constant magnetic field in the long
wavelength approximation will be introduced (see \cite{rev}). The
Hamiltonian in the two inequivalent corners of the Brillouin zones can be
put in a compact notation as%
\begin{equation}
H_{0}^{(\lambda )}=v_{F}(\lambda \sigma _{x}p_{x}+\sigma _{y}p_{y})+\left( 
\begin{array}{cc}
\Sigma _{A}^{(\lambda )} & 0 \\ 
0 & \Sigma _{B}^{(\lambda )}%
\end{array}%
\right)  \label{a1}
\end{equation}%
where $\lambda =1$ is for the $K$ valley and $\lambda =-1$ for the $%
K^{\prime }$ valley. A diagonal energy matrix has been introduced for
further application to single-site approximation (CPA). The quasiparticle
momentum is $\mathbf{p-}e\mathbf{A}$, where $e$ is the electron charge and $%
\mathbf{A}$ is the vector potential which in the Landau gauge reads $\mathbf{%
A=}(-By,0,0)$ and $v_{F}=10^{6}m/s$ is the Fermi velocity.\footnote{%
Spin Zeeman energy are completely neglected because the spin splitting is
much smaller than Landau-level separations.} The Hamiltonian of last
equation is invariant under traslations in the $x$ direction which means
that the wave functions can be written as $\psi ^{(\lambda )}=e^{ikx}\left(
\psi _{A}^{(\lambda )},\psi _{B}^{(\lambda )}\right) $. A coordinate
transformation $\overline{y}=\hbar v_{F}k_{x}-eBy$ can be applied followed
by a scale transformation $\overline{y}=\frac{1}{\sqrt{e\hbar B}}\overline{%
\overline{y}}$, finally, introducing the annihilation and creation operators 
$a=\frac{1}{\sqrt{2}}\left( \overline{\overline{y}}+\frac{\partial }{%
\partial \overline{\overline{y}}}\right) $ and $a^{\dag }=\frac{1}{\sqrt{2}}%
\left( \overline{\overline{y}}-\frac{\partial }{\partial \overline{\overline{%
y}}}\right) $, the Hamiltonian of eq.(\ref{intro1}) reads%
\begin{equation}
H^{(\lambda =1)}=\left( 
\begin{array}{cc}
\Sigma _{A}^{(1)} & \gamma a \\ 
\gamma a^{\dag } & \Sigma _{B}^{(1)}%
\end{array}%
\right) \text{ \ \ \ \ \ \ \ }H^{(\lambda =-1)}=\left( 
\begin{array}{cc}
\Sigma _{A}^{(-1)} & -\gamma a^{\dag } \\ 
-\gamma a & \Sigma _{B}^{(-1)}%
\end{array}%
\right)  \label{intro1}
\end{equation}%
where $\gamma =v_{F}\sqrt{2e\hbar B}$. The eigenfunctions and eigenvectors
for the Hamiltonian of last equation reads (see \cite{jsa1} and \cite{jsa2})%
\begin{equation}
\psi _{(n,s,k)}^{(\lambda )}(r)=e^{ikx}\frac{C_{n}}{\sqrt{2L}}\varphi
_{(n,s,k)}^{(\lambda )}(\xi )  \label{intro2}
\end{equation}%
where $\varphi _{(n,s,k)}^{(\lambda )}(\xi )$ reads%
\begin{equation}
\varphi _{(n,s,k)}^{(1)}(\xi )=\left( 
\begin{array}{c}
\alpha _{n}^{(1,s)}\phi _{n-1,k}(\xi )(1-\delta _{n,0}) \\ 
\phi _{n,k}(\xi )%
\end{array}%
\right) \text{ \ \ \ \ \ \ \ \ \ }\varphi _{(n,s,k)}^{(-1)}(\xi )=\left( 
\begin{array}{c}
\phi _{n,k}(\xi ) \\ 
\alpha _{n}^{(-1,s)}\phi _{n-1,k}(\xi )(1-\delta _{n,0})%
\end{array}%
\right)  \label{intro3}
\end{equation}%
being $\phi _{n,k}(\xi )$ the wave function of the harmonic oscillator%
\footnote{%
The factor $(1-\delta _{n,0})\ $is introduced to discriminate the wave
function with $n=0$. In this case, only one sublattice contributes in both
valleys $K$ and $K^{\prime }$.
\par
{}}%
\begin{equation}
\phi _{n,k}(\xi )=\frac{\pi ^{-1/4}}{\sqrt{2^{n}n!}}e^{-\frac{1}{2}\xi
^{2}}H_{n,k}(\xi )  \label{intro4}
\end{equation}%
$\xi =\frac{\overline{\overline{y}}}{l_{B}}-l_{B}k$, $L=\sqrt{A}$ where $A~$%
is the area of the graphene sheet and%
\begin{equation}
\alpha _{n}^{(\lambda ,s)}=\frac{\lambda (\Sigma _{A}^{(\lambda )}-\Sigma
_{B}^{(\lambda )})-s\sqrt{(\Sigma _{A}^{(\lambda )}-\Sigma _{B}^{(\lambda
)})^{2}+4\gamma ^{2}n}}{2\gamma \sqrt{n}}  \label{intro4.00}
\end{equation}%
where $s=\pm 1$ is the conduction (valence) band index. The coefficient $%
C_{n}$ is $C_{n}=\frac{1}{\sqrt{2-\delta _{n,0}}}$ and $l_{B}=\sqrt{\hbar /eB%
}$ is the magnetic length. The eigenvalues of the Hamiltonian reads%
\begin{equation}
E_{n}^{(s)}=\frac{\Sigma _{A}^{(\lambda )}+\Sigma _{B}^{(\lambda )}-s\sqrt{%
(\Sigma _{A}^{(\lambda )}-\Sigma _{B}^{(\lambda )})^{2}+4\gamma ^{2}n}}{2}
\label{intro5}
\end{equation}%
The degeneracy of each level is $g=L^{2}B/\phi _{0}$ where $\phi _{0}=\hbar
/e$ is the quantum of flux. The low energy description is only valid as long
as the characteristic energy of the excitations is not larger than an energy
cutoff $E_{n,1}<E_{C}$, where $\ E_{C}=\hbar v_{F}k_{\Delta }$ and $%
k_{\Delta }~$is a momentum cutoff. We can choose $k_{\Delta }$ in such a way
to conserve the total number of states in the Brillouin zone, that is, $\pi
k_{\Delta }^{2}=(2\pi )^{2}/A_{C}$, where $A_{C}=3\sqrt{3}a^{2}/2$ is thea
area of the hexagonal lattice (see \cite{peres-guinea}). Then, using eq.(\ref%
{intro5}), $E<E_{C}$ implies that $n<\frac{1}{2e\hbar B}\left[ \frac{\delta
^{2}}{a}-\frac{\delta (\Sigma _{A}^{(\lambda )}+\Sigma _{B}^{(\lambda )})}{%
av_{F}}+\frac{\Sigma _{A}\Sigma _{B}}{v_{F}^{2}}\right] $ where $\delta =%
\sqrt{8\pi /3\sqrt{3}}$, then for weak magnetic fields, the cutoff tends to
infinity and for high magnetic fields, the cutoff tends to zero. Eq. (\ref%
{intro5}) indicates that asymmetry in the substitutional impurity energies
opens an energy gap $\Delta _{n}^{(\lambda )}=\sqrt{(\Sigma _{A}^{(\lambda
)}-\Sigma _{B}^{(\lambda )})^{2}+4\gamma ^{2}n}$.

Using the spectral representation, the Green function of this system reads%
\begin{gather}
g_{ij}^{(\lambda )}(r,r^{\prime },E)=\frac{1}{2L}\sum\limits_{n=0}^{+n_{%
\Delta }}\sum\limits_{s=\pm 1}^{{}}\int_{-\infty }^{+\infty }
\label{intro5.1} \\
\frac{(\alpha _{n}^{(\lambda ,s)})^{\left\vert \lambda +1-i-j\right\vert
}e^{ik(x-x^{\prime })}\phi _{n-i+\frac{1-\lambda }{2},k}^{\ast }(\xi
^{\prime })\phi _{n-j+\frac{1-\lambda }{2},k}(\xi )(1-\delta
_{n,0})^{\left\vert \lambda +1-i-j\right\vert }dk}{(2-\delta _{n,0})\left(
E-E_{n}^{(s)}\right) }  \notag
\end{gather}%
where for the moment $i$, $j=0,1\,\ $where $0$($1$)$~$represent the
sublattice $A$($B$). We can perform the integration in $k$ by using eq.(\ref%
{intro4}) and completing squares 
\begin{equation}
e^{ik(x-x^{\prime })}e^{-\frac{1}{2}(\frac{\overline{\overline{y}}}{l_{B}}%
-l_{B}k)^{2}}e^{-\frac{1}{2}(\frac{\overline{\overline{y}}^{\prime }}{l_{B}}%
-l_{B}k)^{2}}=e^{-l_{B}^{2}(k-\frac{\overline{\overline{y}}+\overline{%
\overline{y}}^{\prime }+i(x-x^{\prime })}{2l_{B}^{2}})^{2}}e^{-\frac{1}{%
4l_{B}^{2}}\left[ (\overline{\overline{y}}-\overline{\overline{y}}^{\prime
})^{2}+(x-x^{\prime })^{2}\right] +\frac{i}{2l_{B}^{2}}(\overline{\overline{y%
}}+\overline{\overline{y}}^{\prime })(x-x^{\prime })}  \label{intro8}
\end{equation}%
Introducing the following coordinate transformation%
\begin{equation}
q=-l_{B}k+\frac{\overline{\overline{y}}+\overline{\overline{y}}^{\prime
}+i(x-x^{\prime })}{2l_{B}}  \label{intro9}
\end{equation}%
and using the following relation (see eq.(7.377$^{8}$) of page 804 of \cite%
{laguerre}) the Green function reads%
\begin{gather}
g_{ij}^{(\lambda )}(r,r^{\prime },E)=\frac{(-1)^{j-i}f(r,r^{\prime
})^{j-i}e^{-\frac{\rho ^{2}}{2}+i\eta (r,r^{\prime })}}{2L}\times
\label{intro11} \\
\sum\limits_{n=0}^{+n_{\Delta }}\frac{(1-\delta _{n,0})^{\left\vert \lambda
+1-i-j\right\vert }}{(2-\delta _{n,0})}T_{n}^{(\lambda ,i,j)}\sqrt{\frac{%
2^{j-i}(n-j+\frac{1-\lambda }{2})!}{(n-i+\frac{1-\lambda }{2})!}}L_{n-j+%
\frac{1-\lambda }{2}}^{\left\vert j-i\right\vert }(\rho ^{2})  \notag
\end{gather}%
where the sum in $s$ has been performed and where 
\begin{equation}
f(r,r^{\prime })=\frac{\overline{\overline{y}}-\overline{\overline{y}}%
^{\prime }-i(x-x^{\prime })}{2l_{B}}  \label{intro12}
\end{equation}%
$\eta (r,r^{\prime })$ is a gauge term that reads%
\begin{equation}
\eta (r,r^{\prime })=\frac{1}{2l_{B}^{2}}(\overline{\overline{y}}+\overline{%
\overline{y}}^{\prime })(x-x^{\prime })  \label{intro13}
\end{equation}%
and%
\begin{equation}
\rho ^{2}=\frac{\left\vert r-r^{\prime }\right\vert ^{2}}{2l_{B}^{2}}
\label{intro13.1}
\end{equation}%
The $T_{n}^{(\lambda ,i,j)}$ matrix elements read\footnote{%
The $T_{n}^{(\lambda ,i,j)}$ coefficients have been introduced to write eq.(%
\ref{intro11}) in a more compact way, although its units are $E^{-1}$ which
are the units of the Green function.}%
\begin{gather}
T_{n}^{(\left\vert \lambda \right\vert ,0,0)}=T_{n}^{(-\left\vert \lambda
\right\vert ,1,1)}=\frac{-2En\gamma ^{2}+(2n\gamma ^{2}+EQ_{1}^{(\lambda
)}-Q_{2}^{(\lambda )})Q_{1}^{(\lambda )}-Q_{2}^{(\lambda )}E+Q_{3}^{(\lambda
)}}{n\gamma ^{2}[-E^{2}+n\gamma ^{2}+\Sigma _{A}^{(\lambda )}(E-\Sigma
_{B}^{(\lambda )})+E\Sigma _{B}^{(\lambda )}]}  \label{intro11.0} \\
T_{n}^{(\lambda ,1,0)}=T_{n}^{(\lambda ,0,1)}=\frac{-2n\gamma ^{2}-\lambda
(\Sigma _{A}^{(\lambda )}-\Sigma _{B}^{(\lambda )})[E+Q_{1}^{(\lambda
)}-\Sigma _{B}^{(\lambda )}-\Sigma _{A}^{(\lambda )}]}{\gamma \sqrt{n}%
[n\gamma ^{2}-E^{2}+\Sigma _{A}^{(\lambda )}(E-\Sigma _{B}^{(\lambda
)})+E\Sigma _{B}^{(\lambda )}]}  \notag \\
T_{n}^{(1,1,1)}=T_{n}^{(-1,0,0)}=\frac{-2E+\Sigma _{A}^{(\lambda )}+\Sigma
_{B}^{(\lambda )}}{n\gamma ^{2}-E^{2}+\Sigma _{A}^{(\lambda )}(E-\Sigma
_{B}^{(\lambda )})+E\Sigma _{B}^{(\lambda )}}  \notag
\end{gather}%
where%
\begin{equation}
Q_{i}^{(\lambda )}=\frac{1}{2}[(1+\lambda )(\Sigma _{A}^{(\lambda
)})^{i}-(\lambda -1)(\Sigma _{B}^{(\lambda )})^{i}]  \label{intro11.1}
\end{equation}%
in the case that $\Sigma _{A}^{(\lambda )}=\Sigma _{B}^{(\lambda )}=0$, eq.(%
\ref{intro11}) is identical to eq.(11) and eq.(12) for $\lambda =1$ of \cite%
{rusin}. It should be pointed out that, as it was shown in eq.(\ref{intro3})
the application of a magnetic field to graphene allows to obtain different
wavefunctions for sublattice $A$ and $B$ in each valley. This is different
of what occurs when $B$ is zero in the effective low-energy description,
because in this case the wavefunctions in each sublattice site differ only
by a phase which depends on the polar angle of the wave vector $\mathbf{k}$.
For this, the probability amplitudes in both sublattice sites are identical.
This crucial difference is of major importance when diagonal disorder is
introduced because it will increase the availability of states near the
Fermi energy without disabling the sublattice asymmetry introduced by the
magnetic field.

\section{Single-site approximation}

To apply CPA\ we can introduce impurity potentials in the $A$ and $B$
sublattices in the following form%
\begin{equation}
H=v_{F}(\lambda \sigma _{x}p_{x}+\sigma _{y}p_{y})+\left( 
\begin{array}{cc}
V_{A} & 0 \\ 
0 & V_{B}%
\end{array}%
\right)  \label{cpa1}
\end{equation}%
where the potential $V^{A/B}$ reads%
\begin{equation}
V^{A/B}=\underset{r_{i}}{\overset{N_{A/B}}{\sum }}V_{i}^{A/B}\left\vert
r_{i}\right\rangle \left\langle r_{i}\right\vert  \label{cpa2}
\end{equation}%
and $V_{i}^{A/B}$ are the on-site impurity energies with a probability
distribution $P(V_{i}^{A/B})$ which is the same for both sublattices and
pseudospin valleys.\footnote{%
The random distribution is identical for $A$ and $B$ sublattices.} To apply
CPA, we can introduce an effective Hamiltonian%
\begin{equation}
H_{ef}=v_{F}(\lambda \sigma _{x}p_{x}+\sigma _{y}p_{y})+\Sigma ^{(\lambda
)}(z)  \label{cpa3}
\end{equation}%
where $\Sigma (z)$ is a energy-dependent self-energy matrix which reads%
\begin{equation}
\Sigma ^{(\lambda )}(z)=\left( 
\begin{array}{cc}
\Sigma _{A}^{(\lambda )}(z) & 0 \\ 
0 & \Sigma _{B}^{(\lambda )}(z)%
\end{array}%
\right)  \label{cpa3.1}
\end{equation}%
Then, an impurity can be introduced in the effective medium in an specific
site $r_{0}$ by defining the following Hamiltonian%
\begin{equation}
H_{h}=H_{ef}+W\left\vert r_{0}\right\rangle \left\langle r_{0}\right\vert
\label{cpa4}
\end{equation}%
where $W$ reads%
\begin{equation}
W=\left( 
\begin{array}{cc}
V_{A}-\Sigma _{A}(z) & 0 \\ 
0 & V_{B}-\Sigma _{B}(z)%
\end{array}%
\right)  \label{cpa4.0}
\end{equation}%
Applying the self-consistency condition for the Green function $%
G_{ef}=\left\langle G_{h}\right\rangle $, where $\left\langle
...\right\rangle $ means configurational averaging, we obtain the Soven
equation in matrix form\footnote{%
For more details about CPA\ see \cite{gon2}, chapter VIII.}%
\begin{equation}
\left\langle W\left[ I-g_{ef}(r_{0},r_{0})W\right] ^{-1}\right\rangle =0
\label{cpa4.1}
\end{equation}%
where $g_{ef}$ is the Green function matrix with magnetic field associated
to the effective Hamiltonian of eq.(\ref{cpa3}), which is identical to eq.(%
\ref{a1}) and implies that the effective Green function is that of eq.(\ref%
{intro11}), which must be evaluated at $r_{0}$ in both coordinates arguments 
\begin{equation}
g_{ij}^{(\lambda )}(r_{0},r_{0},E)=-\delta _{ij}\frac{1}{2L}%
\sum\limits_{n=0}^{+n_{\Delta }}\frac{(1-\delta _{n,0})^{\left\vert \lambda
+1-i-j\right\vert }}{(2-\delta _{n,0})}T_{n}^{(\lambda ,i,j)}  \label{cpa6}
\end{equation}%
The non-diagonal elements are zero due to the factor $f(r,r)=0$ for
identical points. This result implies that eq.(\ref{cpa4.1}) contains a
system of two coupled self-energies that must be solved self-consistently.
Applying the configurational averaging by using the probability distribution
for random binary alloy $P(V_{i})=c\delta (V-\epsilon _{1})+(1-c)\delta
(V-\epsilon _{2})$, the system of eq.(\ref{cpa4.1}) reads%
\begin{eqnarray}
\frac{c(\epsilon _{1}-\Sigma _{A}^{(\lambda )})}{1-g_{AA}^{(\lambda )}\left[
\epsilon _{1}-\Sigma _{A}^{(\lambda )}\right] }+\frac{(1-c)(\epsilon
_{2}-\Sigma _{A}^{(\lambda )})}{1-g_{AA}^{(\lambda )}\left[ \epsilon
_{2}-\Sigma _{A}^{(\lambda )}\right] } &=&0  \label{cpa7} \\
\frac{c(\epsilon _{1}-\Sigma _{B}^{(\lambda )})}{1-g_{BB}^{(\lambda )}\left[
\epsilon _{1}-\Sigma _{B}^{(\lambda )}\right] }+\frac{(1-c)(\epsilon
_{2}-\Sigma _{B}^{(\lambda )})}{1-g_{BB}^{(\lambda )}\left[ \epsilon
_{2}-\Sigma _{B}^{(\lambda )}\right] } &=&0  \notag
\end{eqnarray}%
where we have restored the original notation $g_{AA}^{(\lambda
)}=g_{00}^{(\lambda )}$ and $g_{BB}^{(\lambda )}=g_{11}^{(\lambda )}$, where
these functions depend on $\Sigma _{A}^{(\lambda )}$ and $\Sigma
_{B}^{(\lambda )}$ and $c$ is the concentration of $\epsilon _{1}$
impurities that can be located in $A$ or $B$ sublattices and $\epsilon _{2}$
is the on-site impurity energy with probability $1-c$ that can be located in
the $A$ or $B$ sublattice.

\section{Results and discussion}

To apply CPA\ to eq.(\ref{cpa7}) we have to take into account the diagonal
elements of the Green function, which can be solved exactly%
\begin{gather}
g_{AA}^{(1)}(r,r,E)=g_{BB}^{(-1)}(r_{0},r_{0},E)=  \label{r1} \\
-\frac{1}{4L}A^{(\lambda )}\left[ \psi ^{(0)}(n_{\Delta }+1+C^{(\lambda
)})-\psi ^{(0)}(1+C^{(\lambda )})\right]  \notag \\
-\frac{1}{4L}\frac{B^{(\lambda )}}{C^{(\lambda )}}\left[ \psi
^{(0)}(1+C^{(\lambda )})+\psi ^{(0)}(n_{\Delta }+1)-\psi ^{(0)}(C^{(\lambda
)}+n_{\Delta }+1)+\gamma _{0}\right]  \notag
\end{gather}%
where $\gamma _{0}$ is the Euler-Mascheroni constant,\thinspace $\psi ^{(0)}$
is the Digamma function and the coefficients $A$, $B$ and $C$ reads%
\begin{gather}
A^{(\lambda )}(E)=2\gamma ^{-2}\left[ Q_{1}^{(\lambda )}-E\right]  \label{r2}
\\
B^{(\lambda )}(E)=\gamma ^{-4}\left[ (EQ_{1}^{(\lambda )}-Q_{2}^{(\lambda
)})Q_{1}^{(\lambda )}-Q_{2}^{(\lambda )}E+Q_{3}^{(\lambda )}\right]  \notag
\\
C^{(\lambda )}(E)=\gamma ^{-2}\left[ -E^{2}+\Sigma _{A}^{(\lambda
)}(E-\Sigma _{B}^{(\lambda )})+E\Sigma _{B}^{(\lambda )}\right]  \notag
\end{gather}%
In the same way%
\begin{gather}
g_{BB}^{(1)}(r,r,E)=g_{AA}^{(-1)}(r,r,E)=  \label{r3} \\
-\frac{1}{2L}\frac{-2E+\Sigma _{A}^{(\lambda )}+\Sigma _{B}^{(\lambda )}}{%
-E^{2}+\Sigma _{A}^{(\lambda )}(E-\Sigma _{B}^{(\lambda )})+E\Sigma
_{B}^{(\lambda )}}  \notag \\
-\frac{1}{4L}D^{(\lambda )}\left[ \psi ^{(0)}(F^{(\lambda )}+n_{\Delta
}+1)-\psi ^{(0)}(1+F^{(\lambda )})\right]  \notag
\end{gather}%
where%
\begin{gather}
D^{(\lambda )}(E)=\gamma ^{-2}\left[ -2E+\Sigma _{A}^{(\lambda )}+\Sigma
_{B}^{(\lambda )}\right]  \label{r4} \\
F^{(\lambda )}(E)=\gamma ^{-2}\left[ -E^{2}+\Sigma _{A}^{(\lambda
)}(E-\Sigma _{B}^{(\lambda )})+E\Sigma _{B}^{(\lambda )}\right]  \notag
\end{gather}%
From eq.(\ref{r1}) and eq.(\ref{r3}), the contribution of the conduction
(valence) band in the sublattice $A$ ($B$) equal the valence (conduction)
band in the sublattice $B$ ($A$), which reflects the fact that, although
on-site impurities were introduced, the configurational averaging restore
the full symmetry of the Hamiltonian. A\ convenient way to treat the coupled
system of Soven equations of eq.(\ref{cpa7}) is to consider an iterative
formula equivalent to the CPA\ condition in the following way 
\begin{figure}[tbp]
\centering
\includegraphics[width=130mm,height=80mm]{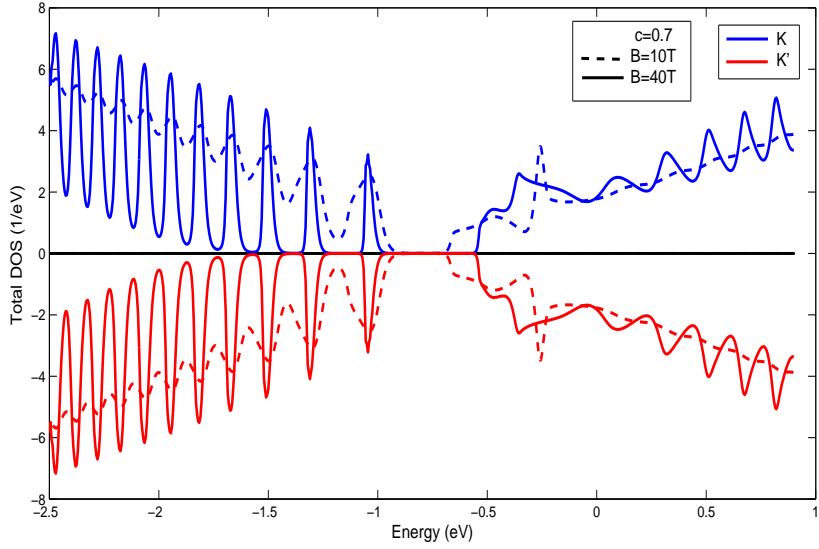}
\caption{Total density of states for two different values of applied
magnetic field. Blue line for $K$ valley and red line for $K^{\prime }$
valley and $\protect\epsilon _{1}=-0.5=-\protect\epsilon _{2}$. }
\label{totaldos}
\end{figure}
\begin{equation}
\Sigma _{A/B}^{(\lambda )}(i+1)=\Sigma _{A/B}^{(\lambda )}(i)+\frac{1}{%
g_{AA/BB}^{(\lambda )}(\Sigma _{A/B}^{(\lambda )}(i),\Sigma _{B/A}^{(\lambda
)}(i))}+\left\langle \frac{1}{\Sigma _{A/B}^{(\lambda )}(i)-V+\frac{1}{%
g_{AA/BB}^{(\lambda )}((\Sigma _{A/B}^{(\lambda )}(i),\Sigma
_{B/A}^{(\lambda )}(i)))}}\right\rangle ^{-1}  \label{r5}
\end{equation}%
\begin{figure}[tbh]
\begin{minipage}{0.48\linewidth}
\includegraphics[width=84mm,height=57mm]{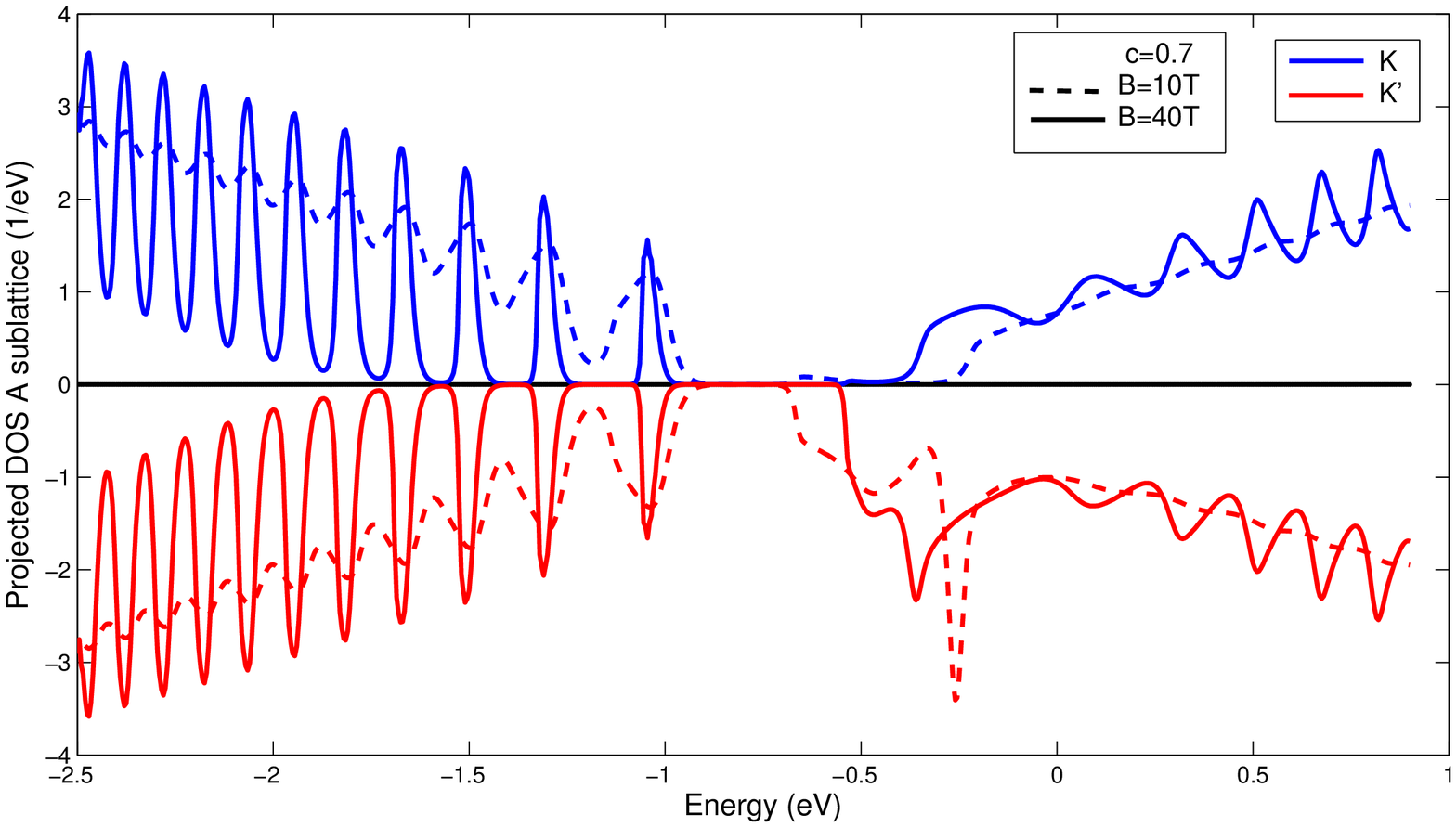} 
\caption{Projected DOS on A sublattice for two different values of magnetic field. Blue line for $K$ valley and red line for $K^{\prime }$ valley and $\epsilon
_{A}=-0.5=-\epsilon _{B}$}
\label{proA}
\end{minipage}
%\quad 
\hspace{0.07cm} 
\begin{minipage}{0.5\linewidth}
\centering
\includegraphics[width=86mm,height=57mm]{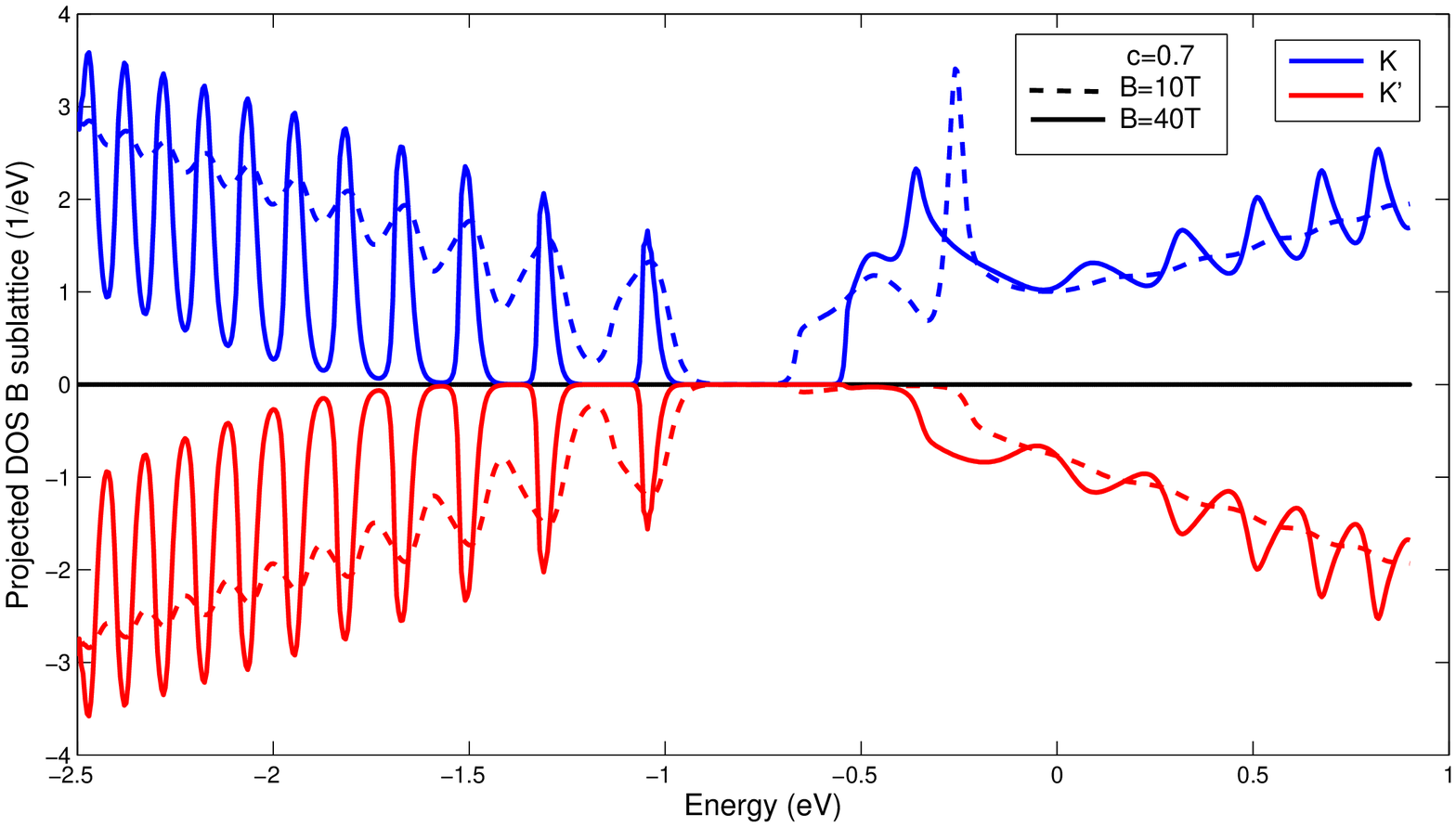}
\caption{Projected DOS on B sublattice for two different values of magnetic field. Blue line for $K$ valley and red line for $K^{\prime }$ valley and $\epsilon
_{A}=-0.5=-\epsilon _{B}$}
\label{proB}
\end{minipage}
%\label{pdosA} \label{pdos}
%\label{pro}
\end{figure}
In order to obtain the value of the self-energies, the iteration described
above for both sublattices is repeated until $\left\vert \Sigma
_{A/B}^{(\lambda )}(i+1)-\Sigma _{A/B}^{(\lambda )}(i)\right\vert
<\varepsilon $, where $\varepsilon $ determines the requested precision of
the calculation. The total and partial density of states can be computed as%
\begin{equation}
\rho (E)=-\frac{1}{\pi }\text{Im}g(E-\Sigma (E))  \label{cpa1.7}
\end{equation}%
\begin{figure}[tbp]
%\centering
\begin{minipage}[b]{0.53\linewidth}
\includegraphics[width=90mm,height=60mm]{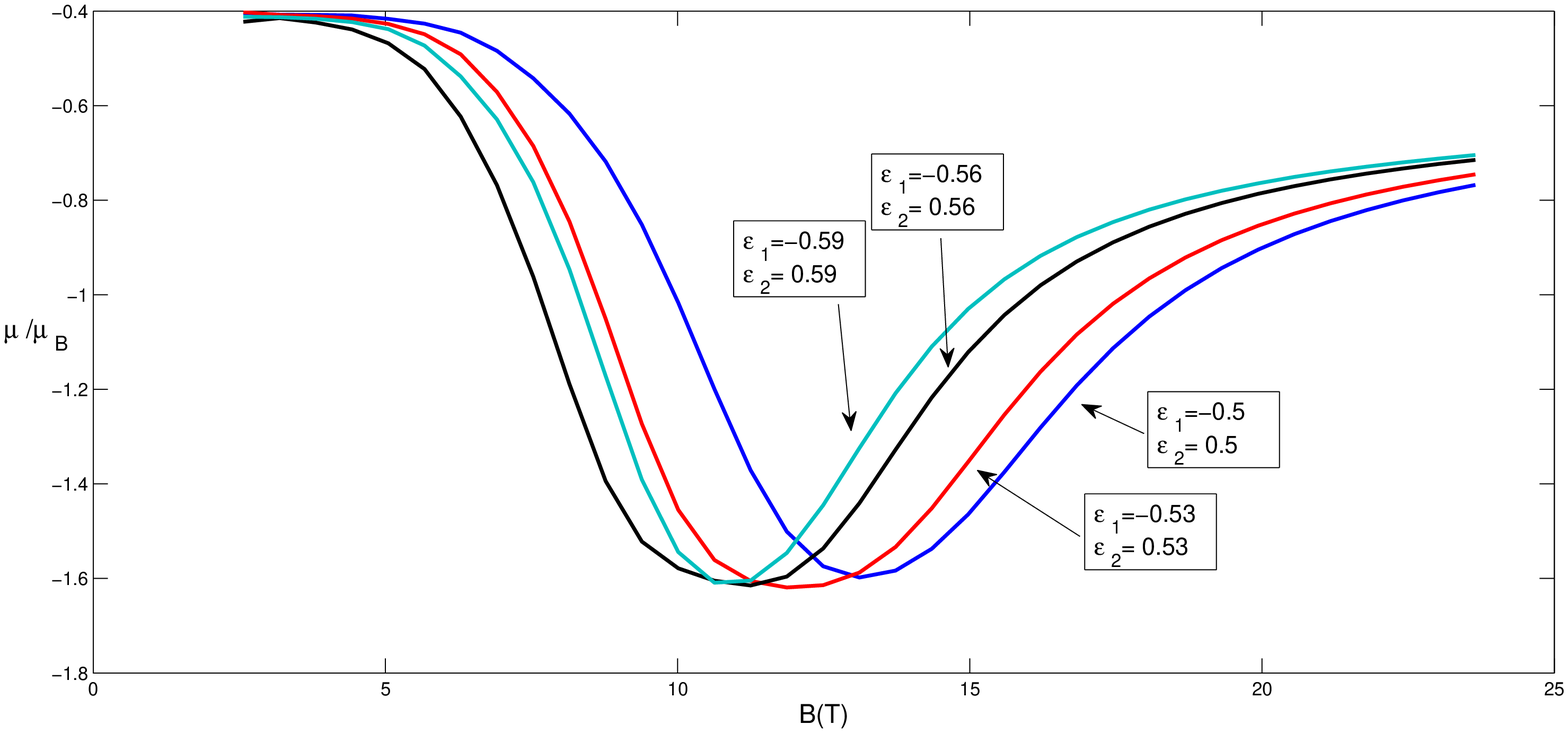}
\caption{Magnetization of A sublattice due to relative difference between density of state contributions for the two valleys pseudospin.}
\label{magA}
\end{minipage}
%\quad
\begin{minipage}[b]{0.53\linewidth}
\includegraphics[width=90mm,height=60mm]{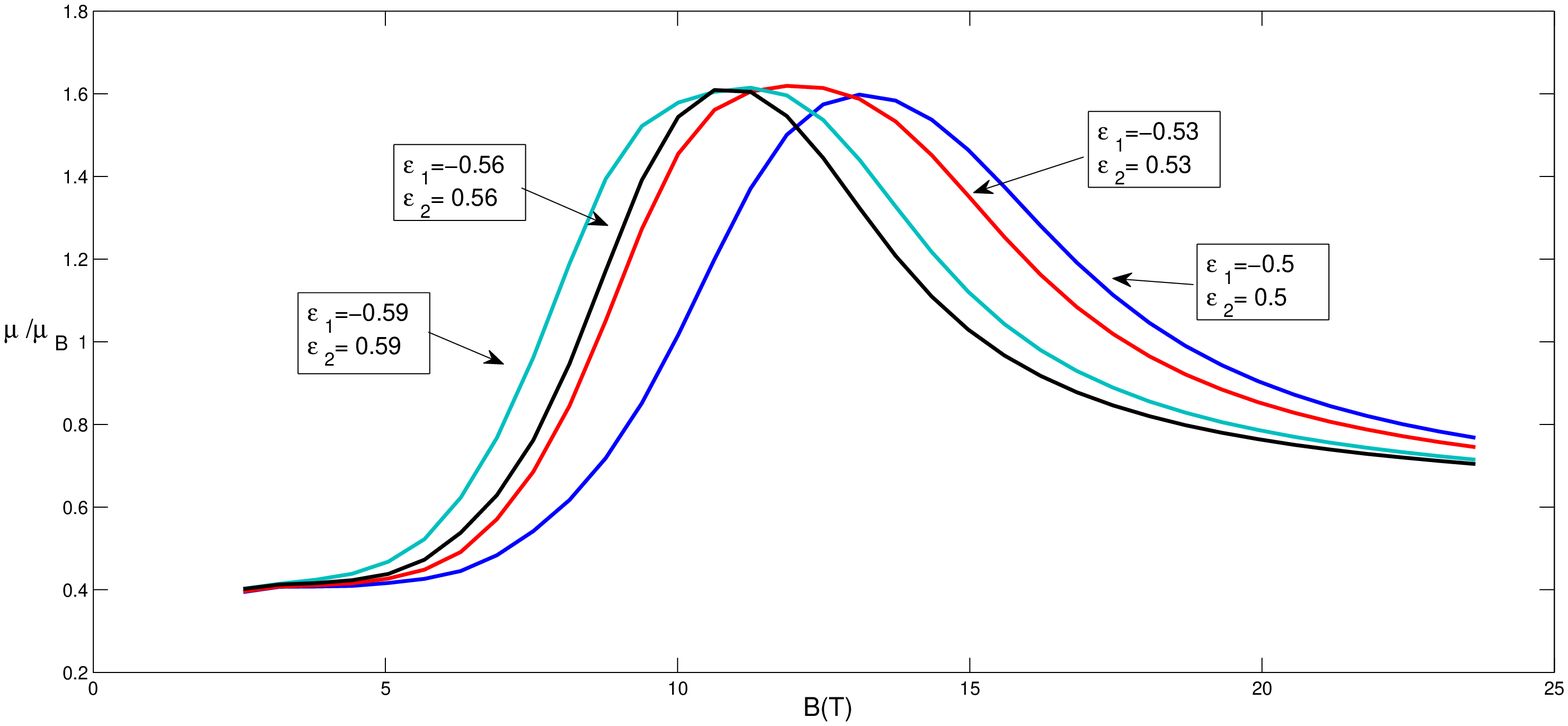}
\caption{Magnetization of B sublattice due to relative difference between density of state contributions for the two valleys pseudospin.}
\label{magB}
\end{minipage}
\label{pdosB} \label{magn}
\end{figure}
In figure \ref{totaldos} and figures \ref{proA} and \ref{proB} the total
density of states and projected DOS in $A$ and $B$ sublattice is shown for
both pseudospin valleys using $\epsilon _{1}=-0.5=-\epsilon _{2}$, $c=0.7$
and for $B=10T$ and $B=40T$. For the specific concentration chosen, we are
considering acceptor impurities. The total DOS\ shows effectively that no
broken pseudospin symmetry has been produced by averaging over disorder, but
figure \ref{proA} and \ref{proB} shows an asymmetry in the $K$ and $%
K^{\prime }$ valleys for the projected density of states below the Fermi
level in both sublattices. There is an enhancement of the population for $A$
sublattice in the $K$ valley with respect to the $K^{\prime }$ valley and
the opposite for $B$ sublattice. This behavior is expected due to the
asymmetry of the lowest Landau level in both sublattices, where the shifted
peak correspond to the impurity contribution. The behavior of the density of
states near the energy threshold where the approximation holds is expected
and reflects the linear regime with the peaks corresponding to the
broadening of Landau levels (see figure 10 of \cite{peres-guinea}). In turn,
the impurities open an energy gap at the Dirac point (see \cite{koshi})\
which lies below the Fermi level for both magnetic field strength, although
for higher $B$, the impurity band is reduced and shifted away the Fermi
level. In the projected DOS figures, the energy gap introduced by the
diagonal impurities is different for $K$ and $K^{\prime }$ valley, which is
a consequence of the self-energy asymmetry results obtained from CPA. In the
case that $c=0.5$, the asymmetry in the gap for $K$ and $K^{\prime }$
valleys should be of major importance because in this case, the impurity
band will be located in the Fermi level and for half-filling, a net
magnetization will appears.\footnote{%
The $c=0.5$ impurity concentration has not been studied in this work because
convergence for self-energy in CPA is difficult to achieve in this case.} 
\begin{figure}[tbp]
\centering
\includegraphics[width=130mm,height=80mm]{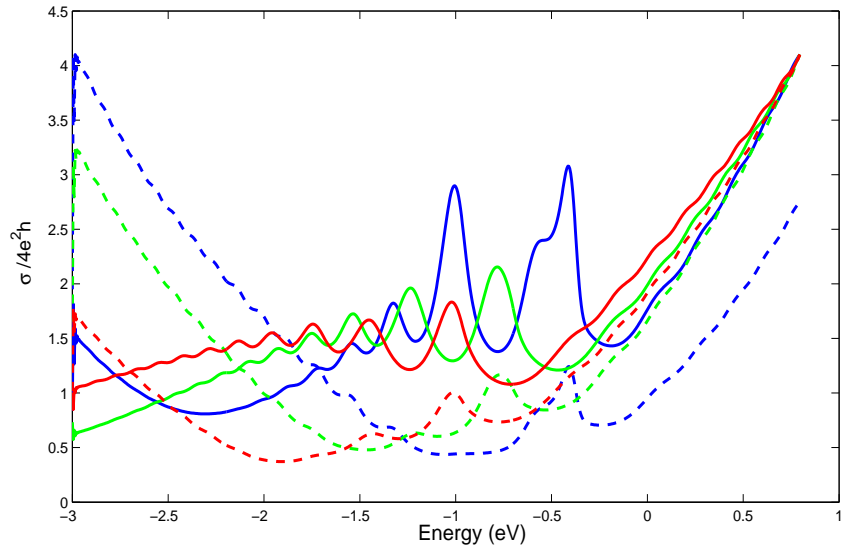}
\caption{Conductivity as a function of energy for different on-site
impurities. Blue line $\protect\epsilon _{1}=\protect\epsilon _{2}=-0.6$,
green line $\protect\epsilon _{1}=\protect\epsilon _{2}=-0.8$ and red line
for $\protect\epsilon _{1}=\protect\epsilon _{2}=-1$. Dashed line for
sublattice $A$ scattering contribution to conductivity. The magnetic field
strength used is $B=10T$.}
\label{conduc}
\end{figure}
In figure \ref{magA} and \ref{magB}, the net magnetization in both
sublattices is shown for different impurity on-site energies where%
\begin{equation}
\mu _{A/B}/\mu _{0}=\int_{-E_{C}}^{E_{F}}[\rho _{A/B}^{K}(E)-\rho
_{A/B}^{K^{\prime }}(E)]dE  \label{cpa1.8}
\end{equation}%
where $\mu _{0}$ is the Bohr magneton. As it is expected, both magnetization
are equal an opposite, which implies that there is no net magnetization when
both sublattices are taking into account. 
\begin{figure}[tbp]
\centering
\includegraphics[width=130mm,height=80mm]{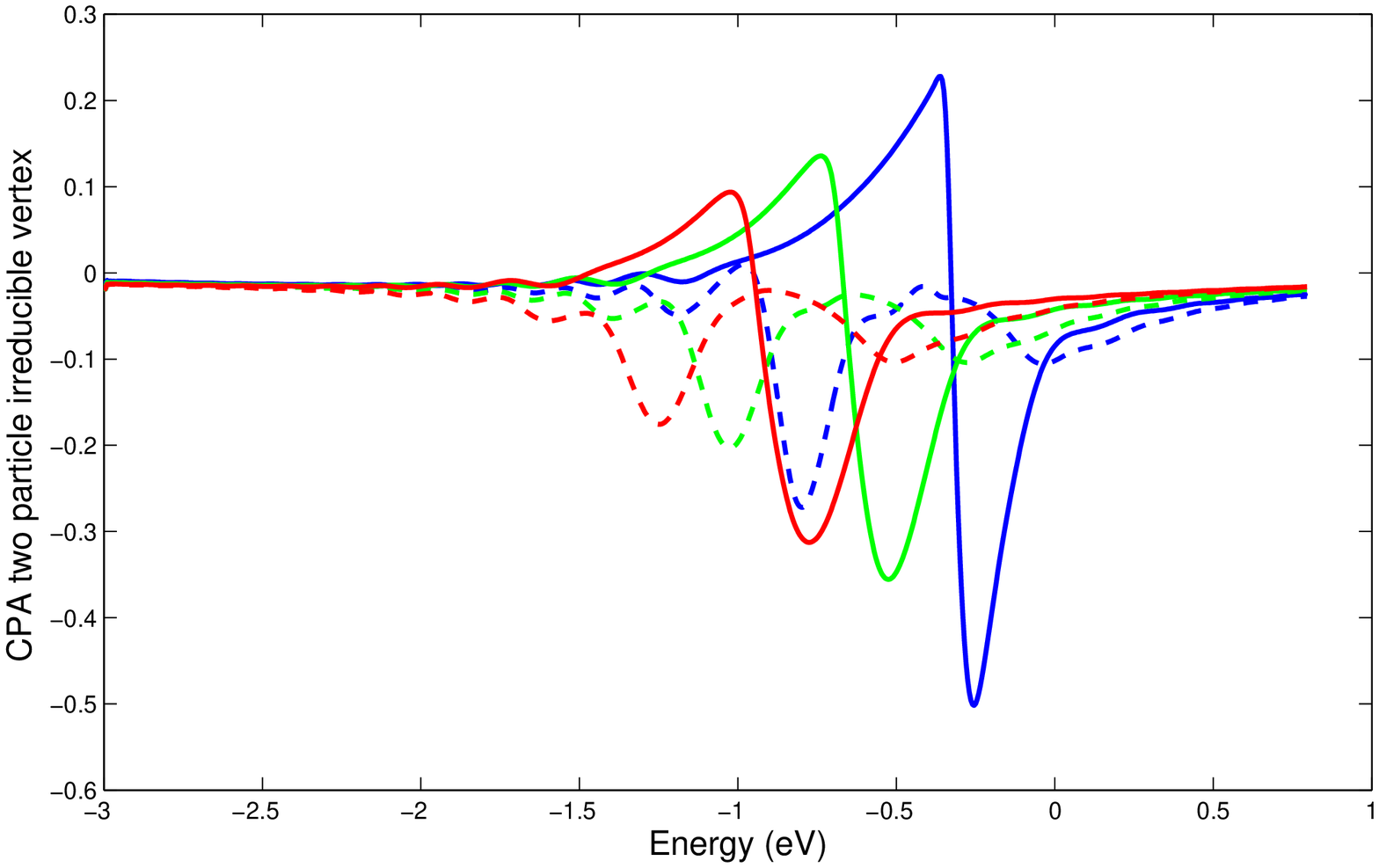}
\caption{Vertex function as a function of energy for different on-site
impurities. Blue line $\protect\epsilon _{1}=\protect\epsilon _{2}=-0.6$,
green line $\protect\epsilon _{1}=\protect\epsilon _{2}=-0.8$ and red line
for $\protect\epsilon _{1}=\protect\epsilon _{2}=-1$. Thick line for
sublattice $A$ scattering and dashed line for sublattice $B$ scattering
contribution to conductivity. The magnetic field strength used is $B=10T$.}
\label{vertex}
\end{figure}
In both sublattices, the magnetization present a maximum value that
decreases with an increasing on-site impurity potentials, which implies that
it is always possible to maximize magnetization with an specific choice of
the parameters. The shift of the maximum with the displacement of impurity
on-site energy is the combination of two physical process: the enhanced
asymmetry between $K$ and $K^{\prime }$ density of states near the Fermi
level when the strength of the impurities is raised and the spreadening of
the impurity band due to the weak magnetic field, although when electron
many body effects are taking into account, this behavior should be reduced
because high density of states available from the $K$ or $K^{\prime }$ band
increase the energy caused by spin-alignment which in turn results in a
small contribution to the exchange energy.

In figure \ref{conduc}, the conductivity is plotted against energy using
eq.(14) of \cite{lovt}, where Kubo formalism is used, for different values
of impurity on-site energies. The position of the conductivity peaks
correspond to the minimum and maximum of the energy band, where the density
of states is the largest. In dashed line, the contribution to conductivity
for sublattice $A$ scattering is taking into account. In particular, for
higher values of impurity on-site energies, the contribution to total
conductivity peaks are given by $A$ sublattice scattering (red line), but is
not the case for the blue line and green line, where there is a contribution
from $B$ sublattice scattering for the second peak respectively. In figure %
\ref{vertex} the local two-particle irreducible vertex $\lambda $ consistent
with the coherent potential approximation for the self-energy $\Sigma $ is
plotted against energy, for retarded and advanced Green functions (see \cite%
{kada}, \cite{jindra}, \cite{janis} and \cite{janis2}) with the same
parameters as figure \ref{conduc}. The peak of the vertex for $A$ sublattice
scattering \ (blue thick line in figure \ref{vertex}) is related to peak
near the Fermi level in figure \ref{conduc}, while the second peak is given
by the contribution of vertex for $B$ sublattice scattering (blue dashed
line in figure \ref{vertex}) with its respective decrease in the
conductivity contribution (see blue dashed line in figure \ref{conduc}). In
turn, from figure \ref{vertex}, the vertex function for $A$ sublattice
scattering shows a singular behavior where the retarded and advanced Green
function in the real part of the energy are identical. The same behavior is
not present for the $B$ sublattice scattering which is related to the
different contributions of Landau levels for $A$ and $B$ sublattice. The
singular behavior of the vertex function implies a critical contribution to
conductivity that change sign and can be related to weak anti-localization
(see \cite{alt}, \cite{lar}, \cite{ber}, \cite{kh} and \cite{susurra}).

These results reflect the fact that, although a simple model has been
considered, where the impurities are diagonal in the sublattice basis and
short-ranged and averaging has been applied, the application of a constant
magnetic field introduce subtleties in the valley properties of electrons.
In particular, the different Landau levels contributions to the
eigenfunctions for each sublattice increase the asymmetry in the orbital
magnetic moment present in electrons in the two Dirac points and for
particular values of impurity concentrations and magnetic field strength,
the net magnetization could be larger to create and manipulate polarized
valley currents. From the experimental viewpoint, doping in graphene can be
obtained through electric doping by changing gate voltage (see \cite{novo2}
and \cite{small}) or by chemical doping, which is discussed as surface
transfer doping and subtitutional doping (see \cite{rao} and \cite{wei}).
For a general review of the experimental procedure to obtain doping
asymmetry in graphene see \cite{hont}, \cite{weh} and the references therein
and for the direct experimental determination of the chemical bonding of
individual impurity atoms see \cite{zzz}. In the case that p-type and n-type
doping can be achieved in a controled way and where the experimental
determination of the on-site energies results in slightly different values,
then it would be possible to tune up population difference in both valleys
in each sublattice site through the magnetic field strength and valley
currents could be obtained by applying a gate voltage.

\section{Conclusion}

We have investigated the behavior of the density of states in both
sublattices and for the different valley pseudospins in graphene with
disorder and magnetic field in the low energy effective-mass theory. By
applying coherent potential approximation for a binary random alloy and
considering the contribution for both sublattices on the effective
self-energies, a coupled system of Soven equations can be obtained to be
solved numerically. We have shown that a band gap is opened by an asymmetric
on-site impurity concentration and the graphene electrons acquire an
anomalous magnetic moment, which is opposite in different valleys, similar
to a real spin. The valley contributions to the projected density of states
for each sublattice are not simetrical and the differences depend highly in
the interplay between the impurity band, the band edges and the broadening
of the Landau levels. In the case that spin-orbit interaction and Zeeman
effect are considered, the asymmetry between valleys can be enhanced by
splitting of Landau levels. In turn, conductivity and CPA local
two-irreducible vertex has been computed showing that scattering from
sublattice $A$ and $B$ do not contribute equally, which implies that
graphene Bloch electrons correlations in the long wavelength approximation
in averaged disorder and magnetic field develop unusual behavior for
critical energy values.

\section{Acknowledgment}

This paper was partially supported by grants of CONICET (Argentina National
Research Council) and Universidad Nacional del Sur (UNS) and by ANPCyT
through PICT 1770, and PIP-CONICET Nos. 114-200901-00272 and
114-200901-00068 research grants, as well as by SGCyT-UNS., E.A.G., P.V.J.
and J. S. A. are members of CONICET. P.B. is a fellow researcher at this
institution. We are grateful to the Abdus Salam International Centre for
Theoretical Physics (ICTP) to allow J. S. A. to spend his time in the
Condensed Matter Section and to be benefited from helpul discussions with
the staff.

\section{Author contributions}

All authors contributed equally to all aspects of this work.

\end{document}